# Classical ODE and PDE Which Obey Quantum Dynamics

Paul J. Werbos[*]
National Science Foundation[†], Room 675
Arlington, VA 22203

**ABSTRACT**

It is well known that classical systems governed by ODE or PDE can have extremely complex emergent properties. Many researchers have asked: is it possible that the statistical correlations which emerge over time in classical systems would allow effects as complex as those generated by quantum field theory (QFT)? For example, could parallel computation based on classical statistical correlations in systems based on continuous variables, distributed over space, possibly be as powerful as quantum computing based on entanglement? This paper proves that the answer to this question is essentially "yes," with certain caveats.

More precisely, the paper shows that the statistics of many classical ODE and PDE systems obey dynamics remarkably similar to the Heisenberg dynamics of the corresponding quantum field theory (QFT). It supports Einstein's conjecture that much of quantum mechanics may be derived as a statistical formalism describing the dynamics of classical systems.

Predictions of QFT result from *combining* quantum dynamics with quantum measurement rules. Bell's Theorem experiments which rule out "classical field theory" may therefore be interpreted as ruling out classical assumptions about measurement which were not part of the PDE. *If* quantum measurement rules can be *derived* as a consequence of quantum dynamics and gross thermodynamics, they should apply to a PDE model of reality just as much as they apply to a QFT model. This implies: (1) the real advantage of "quantum computing" lies in the exploitation of quantum measurement effects, which may have possibilities well beyond today's early efforts; (2) Lagrangian PDE models assuming the existence of objective reality should be reconsidered as a "theory of everything." This paper will review the underlying mathematics, prove the basic points, and suggest how a PDE-based approach might someday allow a finite, consistent unified field theory far simpler than superstring theory, the only known alternative to date.

## 1. Introduction and Summary

This paper assumes that the reader is already familiar with classical Lagrangian field theory. However, it will not assume a familiarity with QFT.

Section 2 will review basic concepts of QFT, and explain the discussion of quantum measurement and computation in the Abstract.

Section 3 will review the elegant mathematics of creation and annihilation operators. There are two types of creation and annihilation operator used in QFT – "bosonic" and "fermionic." Sections 3 through 5 will only discuss bosonic operators and bosonic QFT. The standard model of physics uses a mix of bosonic operators (e.g., to represent light) and fermionic operators (e.g., to represent electrons, protons and neutrons); however, well-known physicists like Coleman[1975], Mandelstam[1975] and

---

Witten [1984] have proven equivalences between fermionic QFTs and bosonic QFTs, as I will discuss in section 6.

Section 3 will also explain the basic equations of Heisenberg dynamics. For second-order Lagrangian ODE systems over $R^n$, the Heisenberg dynamical equations for the corresponding QFT (in the notation of Weinberg[1995, chapter 7]) are:

$$\dot{Q}_j = -i[Q_j, H_n] \tag{1}$$

$$\dot{P}_j = -i[P_j, H_n] \tag{2}$$

where $Q_j$, $P_j$ and $H_n$ are linear operators composed from creation and annihilation operators, and where the square brackets denote the commutator of the two operators. For PDE systems in d+1 dimensions ( d dimensions of space and 1 of time), the equations are

$$\dot{Q}_j(\underline{x}) = -i[Q_j(\underline{x}), H_n] \tag{3}$$

$$\dot{P}_j(\underline{x}) = -i[P_j(\underline{x}), H_n] \tag{4}$$

where $Q_j(\underline{x})$ and $P_j(\underline{x})$ are called "field operators" and are also composed of creation and annihilation operators. Because $Q_j$ is actually a function of space *and* time ($\underline{x}$ and t), we will follow Weinberg in assuming:

$$Q_j(\underline{x}, 0) = \Phi_j(\underline{x}) \tag{5}$$

where $\Phi_j(\underline{x})$ is a simpler field operator (the same as what he calls q). Weinberg actually uses equations 3-5 as *definitions*; from them he proves in effect that:

$$\text{Tr}\left(\rho\left(\frac{d}{dt}\bigg|_{t=0} Q_j(\underline{x}, t)\right)\right) = \frac{d}{dt} Tr(\rho \Phi_j), \tag{6}$$

(and likewise for $P_j$ and $\Pi_j$) where $\rho$ is the density matrix of QFT. From there he argues that one can derive multipoint Green's functions and the rest of QFT (except for assumptions about quantum measurement). Mandelstam [1975] uses equivalence in Heisenberg dynamics as the basis for asserting equivalence between a simple bosonic QFT (the quantum sine-Gordon system) and a simple fermionic system. The operator $H_n$ is called "the normal form Hamiltonian operator."

Section 4 will prove that classical second-order Lagrangian ODE systems *also obey* Heisenberg dynamics, without being quantized, in the following sense. I will assume a Lagrangian of the form:

$$L = \tfrac{1}{2}\sum_{j=1}^{n}(\dot{\varphi}_j \pi_j - \dot{\pi}_j \varphi_j) - H(\underline{\varphi},\underline{\pi}) \tag{7}$$

where H is an analytic function and where the state of the system at time t is defined by the two classical vectors $\underline{\varphi}, \underline{\pi} \in R^n$. I will show how the statistical correlations of the field at any time t can be encoded into an operator $\rho(t)$ which I will call "the classical density matrix." For any statistical ensemble of states, I will first prove that:

$$\text{Tr}(\rho \Phi_j) = <\varphi_j> \tag{8}$$

where "$<\varphi_j>$" is the expectation value of $\varphi_j$, and the expression on the left-hand-side is the standard formula used in modern QFT to calculate the expected value of a measured variable $\varphi_j$. I will then prove, more generally, that:

$$\text{Tr}(\rho g_n(\underline{\Phi},\underline{\Pi})) = <g(\underline{\varphi},\underline{\pi})> \tag{9}$$

where $g_n$ is the "normal form" of the function $g(\underline{\Phi}, \underline{\Pi})$. (In physics, $g_n(\underline{\Phi}, \underline{\Pi})$ is usually written as :$g(\underline{\Phi}, \underline{\Pi})$: or as $N(g(\underline{\Phi}, \underline{\Pi}))$.) If we consider the case where g=H, equation 9 implies that the expected energy levels of a classical ensemble would equal $\text{Tr}(\rho H_n)$, the same energy levels predicting by the corresponding QFT. Finally, I will prove that equation 6 holds for the classical density matrix in the ODE case. Equations 6, 8 and 9 are the key results of this paper.

One might try to extrapolate these conclusions to *any* analytic function g(φ, π) by proving:

$$\text{Tr}(\rho(0)G(t)) =< g(\underline{\varphi}(t),\underline{\pi}(t)) >, \qquad (10)$$

where G(t) is the operator defined by the more general Heisenberg equation

$$\dot{G} = -i[G, H_n] \qquad (11)$$

and the initial condition:

$$G(0) = g_n(\underline{\Phi}, \underline{\Pi}) \qquad (12)$$

But this extrapolation does *not* hold exactly in the general case. The implications are beyond the scope of this paper.

Section 5 will not prove the extension of all these results to the PDE case. However, it will define the classical entropy matrix for the PDE case, and prove the analogs of equations 7 and 8. The PDE case is far more complicated than the ODE case, but the same approach should go through, as will be discussed. Many standard texts have *defined* QFT by (1) representing space as a finite lattice of points or modes of vibration; (2) quantizing the resulting ODE; (3) then taking the limit as the lattice spacing goes to zero and the volume of space to infinity.

Finally, section 6 will discuss possibilities for future research. It will detail a strategy for achieving a consistent unified "theory of everything" based on Lagrangian PDE, consistent with local realism. Completion of this program would require contributions from many researchers, from many branches of mathematics and physics. In the end, only experiments will be able to decide between this kind of unified field theory and the alternative offered by superstring theory. There is reason to hope that a successful resolution of these issues, through future research, could have large applications – perhaps even implications for technology. Circa 1900, it seemed implausible that unknown physics could have large-scale practical applications; but history has shown over and over again that we should be prepared for the possibility of further surprises.

## 2. Quantum Basics and Measurement, Bell's Theorem and Computation

### 2.1. Basics and notation

In traditional bosonic QFT, developed circa 1950, the state of our knowledge at any time t was represented as a complex-valued wave function $\psi$. For example, if we are considering a system of N types of simple particles, moving through d-dimensional space, the "wave function" $\psi(t)$ is actually an infinite collection of the pieces:

$$\psi_0$$
$$\psi_1(i, \underline{x})$$
$$\psi_2(i_1, \underline{x}_1; i_2, \underline{x}_2)$$
$$\ldots \qquad (13)$$
$$\psi_n(i_1, \underline{x}_1; i_2, \underline{x}_2; \ldots; i_n, \underline{x}_n)$$
$$\ldots$$

where the index or "spin" (or "isospin") variables "i" are integers between 1 and N, and where the variables $\underline{x}$ represent locations in d-dimensional space. Each component function, $\psi_n$, is symmetric with respect to an interchange of coordinates

$(i_j, \underline{x}_j) \leftrightarrow (i_k, \underline{x}_k)$. The square norm of the wave function, $|\psi|^2$, is the sum of the square norms of each of the pieces, and is required to equal one. Ignoring the index variables, each piece $\psi_n$ is essentially a complex-valued function over $R^{Nd}$. $\psi$ itself may be viewed as a function defined over a space made up of a zero-dimensional space, concatenated with a d-dimensional space, concatenated with a 2d-dimensional space, and so on, with index-variable dimensions inserted as well. This complicated space is commonly called "Fock space." Functions of finite square norm, defined over Fock space, may be viewed as objects or vectors in "Fock-Hilbert" space. For the wave function shown in equations 13, I will call this Fock space "$F^N(d)$."

Complete textbooks on QFT spend considerable time analyzing how wave functions change under Lorentz transformations. However, that is not necessary here. It would require a paper many times longer than this one. Here I will address the general case of Lagrangian systems, which may be relativistic or not, depending on what Lagrangian one chooses. If one chooses a relativistic PDE system, the resulting statistics will of course be relativistic as well.

Traditional QFT was not a theory about how reality works. The wave function $\psi(t)$ does not describe the state of reality, or even a *probability distribution* for possible states of reality, according to the orthodox Copenhagen interpretation of quantum mechanics. In that view, QFT is simply a *procedure* for predicting the probabilities of measurement outcomes, in what comes out of an experiment, as a function of the initial set-up. The procedure has three steps, essentially:

> (1) Initial knowledge about incoming particles is "encoded" into an incoming wave function, $\psi(t_-)$ for the initial time $t_-$;

(2) The "Schrodinger equation" $\partial_t\psi=iHt$ is used to calculate $\psi(t_+)$ at the final time $t_+$, where we use modern units ($\hbar=c=1$) and where "$\partial_t$" is an abbreviation for $(\partial/\partial t)$.;

(3) The quantum measurement formalism is used to "read out" probabilities for measurement results at time $t_+$, depending on what one chooses to measure.

Step 2 is the step which is called quantum dynamics. Steps 1 and 3 are both aspects of quantum measurement, in the broadest sense. The Hamiltonian operator H could in principle be any linear operator, or even nonlinear; however, existing theories of physics choose H by starting from a classical Lagrangian theory, and then "quantizing" it by following a variety of intuitive procedures.

Decades ago, many physicists suggested that the wave function $\psi$ (or something similar) might actually be a real (though complex-valued) field propagating through space. In this view, the space that we live in is actually Fock space. The three-dimensional world that we see is actually just one of many strands within the larger multiverse. In this view, quantum measurement is not a basic law of the universe, but merely an emergent outcome of quantum dynamics. For example, the classic "many worlds" thesis of Everett [1973] claimed to prove that quantum measurement rules fall out directly from quantum dynamics. Bohm and Hiley [1995] have championed another major re-interpretation of QFT, which leads to the many-worlds viewpoint.

The field of quantum computing can mainly be traced back to early seminal research by David Deutsch of Oxford, based on the Bohmian view of QFT. Early concepts of quantum computing could not work out directly in real hardware, because the wave function ψ turns out to be inadequate -- *empirically* -- as a description of our knowledge about time t, when our knowledge includes knowledge about complex objects like lasers and polarizers and crystals. Modern quantum computing is grounded instead on the dynamics of the *density matrix* ρ, which may be defined, crudely, as:

$$\rho = \sum_{\alpha} \Pr(\alpha) \psi_\alpha \psi_\alpha^H \qquad (14)$$

where Pr(α) is the probability of a "state" α, ψ$_\alpha$ is the wave function of that state viewed as a vector in Fock-Hilbert space, and where the superscript "H" represents the Hermitian conjugate (i.e., the complex conjugate of the transpose). Thus instead of relying on a Schrodinger equation as its foundation, modern work relies on the Liouville equations, equations 3 and 4. For example, see Gershenfeld and Chang [1997] for discussion of how a "pseudopure" state may be encoded into a density matrix.

A caveat is in order here, however. Even though the density matrix is the intellectual foundation of most modern work, density matrix calculations can sometimes be very complex. Therefore, Howard Carmichael [1998,2002] has developed a scheme for reducing the density matrix calculations used in quantum optics to a string of connected wave-function calculations, separated by "jumps;" his scheme, the "quantum trajectory simulation" (QTS) scheme, is the primary workhorse of the optics-based parts of quantum computing. Dowling has stated that this breakthrough was a byproduct of discussions of the foundations of quantum theory which had previously been dismissed by many people as "purely philosophical."

**2.2. Bell's Theorems and the debate about reality**

Most of the early discussions about quantum mechanics versus classical physics were highly dependent on semantics and legalistic debating tactics. For example, many debaters started out by defining "classical physics" as the assumption that electrons are perfect point particles (as assumed in the very old Poisson brackets formalism) -- an assumption much closer to today's standard model than to anything Einstein proposed! Only since 1974 has the story become clearer and more rigorous.

Von Neumann [1955] proved that quantum mechanics (as then understood) could not be reconciled with the standard formulations of classical physics. He did not actually endorse the Copenhagen view that objective reality does not exist. Instead, he stated that quantum mechanics is inconsistent with classical notions about *causality*.

Von Neumann's work was not decisive, because he only proved that quantum mechanics *in its entirety* is inconsistent with classic views about causality. We do not have empirical proof that quantum theory is correct for *all possible experiments*.

The decisive breakthrough came about as the result of some theorems proven by J.S. Bell, which inspired the seminal papers of Clauser, Horne, Shimony and Holt [1969], which were popularized in turn by the seminal book of Bell[1987]. The theorems of Clauser et al prove (in their words) that certain specific predictions of QFT cannot be reconciled with any "local causal hidden-variable theory." These predictions have been verified in at least some experiments, to a high degree of accuracy. (It only takes one valid experiment to prove the point!) They define a "hidden-variable theory" to be any theory of physics which assumes the existence of objective reality. The Copenhagen view of QFT satisfies Bell's Theorem by assuming that objective reality does not exist. They define "local" to mean any theory which excludes action at a distance. The many-worlds theories satisfy Bell's Theorem by assuming action at a distance, in one form or another. The only other choice here is to assume that the classical formulation of "causality" does not hold. That is the choice which I have proposed that we reconsider [Werbos, 1973, 1989]. (Many other physicists have argued for similar ideas, starting from the early work of DeBeauregard, continuing through Leggett, Cramer, Penrose and others, and perhaps even DeBroglie himself, depending on how one interprets his later writings.)

The Bell's Theorems make the following assumption about classical causality. The probability of the state of reality at time t. (i.e., $Pr(\varphi(t_.), \pi(t_.))$) is strictly determined by what comes *into* the experiment in forwards time. Thus if there are any polarizers in the experiment, polarizers which the light will not "see" until a later time $\tau$, then the setting of the polarizers cannot have any effect at all on $Pr(\varphi(t_.), \pi(t_.))$. This concept of "causality" is better called "time-forwards causality."

Some classically-minded physicists have been totally shocked by my suggestion that causal effects might propagate backwards in time as well as forwards. Their initial reaction is sometimes as severe, as absolute and as apriori as the reaction of the medieval Catholic Church when confronted with the possibility that the earth might not be the center of the universe. Defenders of the flat earth idea once argued: "The concept of 'up' and 'down' is hard-wired into our very bones; since we ourselves are the center of the universe -- and we have no chance of understanding the universe on any other basis -- we *must* assume that 'up' and 'down' are universal invariants, representing one universal direction." In the same way, philosophical defenders of time-forwards causality have argued that the notions of 'future' and 'past' are hard-wired into our brains, and that we cannot possibly conceive of any alternative to absolute time-forwards causality at all levels of nature. (But in fact, we *can* conceive of time-symmetric PDE and can even learn to live with them!) Some medieval thinkers would probably have preferred to believe that the material universe is a total illusion, and that there is no objective reality at all, rather than admit that the earth is not the center of that reality; this is logically quite parallel to the combination of the Copenhagen view and the classical concept of time-forwards causality.

If we set aside such apriori philosophical commitments, what does the physics itself tell us here? In effect, Bell's Theorem tells us that we cannot maintain local realism and absolute time-forwards causality at the same time. The concept of absolute time-forwards causality was a product of philosophy, not of the PDE. The Lagrangians used in the standard model of physics are all *symmetric* with respect to time. It should not really surprise us that there is a contradiction between assuming that the universe is governed by time-symmetric PDE, and assuming that flows of causality must operate only forwards in time, even at the most microscopic levels of nature.

If we really think about the possibility that the universe might be governed by (mostly) time-symmetric PDE, the real puzzle lies in explaining why we see any tendency at all towards asymmetry, towards time-forwards causality here on earth. (There have been fascinating discussions by Prigogine[1994] and others about the origins of asymmetry in the larger universe, but they go beyond the scope of this paper.) The answer to this puzzle, here on earth, is straightforward: global time-forwards causality results from boundary conditions in space-time. Without the infusion of light from the sun, the earth would be in a state of maximum entropy, where the statistics forwards in time and the statistics backwards in time would be indistinguishable from each other. But the light of the sun, propagating forwards in time, provides the engine of time-forwards free energy which underlies our ability to perform any experiments at all. The great flow of time-forwards free energy, from the sun to the earth, and from the earth on out to deep space, is like a vast river, driving our everyday experience. Like a river on the surface of the earth, it creates a preferred direction of movement on a large scale.

Yet even the most powerful river does not destroy the directional symmetry of the underlying laws of physics at small distances. Water can actually flow upstream for short distances within a river, depending on local conditions like rocks and whirlpools within the river. It is interesting to ask whether conscious efforts to place the rocks could actually change the direction of flow upstream for larger distances than what we see in the natural scene.

Werbos and Dolmatova [2000, section 2] attempted to analyze the interface between microscopic time symmetry and the macroscopic "river" of free energy more mathematically. They concluded that the usual measurement formalism of QFT is a reasonable second-order approximation to what this picture would imply. (Classical causality is a first-order approximation.) But there is reason to believe that a more accurate approximation may be possible, with empirical consequences.

The original Copenhagen theory of measurement assumed that a "measurement" takes place only when a conscious "observer" chooses to perform an act of measurement. In the famous Schrodinger cat thought-experiment, a cat is placed into a mixed state -- neither dead nor alive -- by putting it in front of a gun triggered by a quantum system;

the cat becomes dead or alive (according to Copenhagen) only after a human observer comes in to look at the cat. Yet the results are different if one assumes that the cat itself is an observer. Many physicists have been troubled by the notion that the physics is different, in principle, depending on whether one believes that cats have souls. In order to avoid treating quantum measurement (including cats) as kind of metaphysical law of the universe, many physicists (like Everett[1973]) have argued that the measurement formalism we really use can be derived as an emergent property of quantum dynamics itself; *if this were true*, then the results of this paper suggest that classical Lagrangian systems are *also* governed by quantum measurement. They would be completely equivalent to QFT, and would satisfy Bell's Theorem. They would therefore have to violate classical time-forwards causality. (Experts in nonlinear systems dynamics should note that this does *not* imply that Lagrangian PDE are ill-posed in time-forwards integration; it means, in effect, that correlations across time within the system -- which dominate the experience of observers living within the system -- go both ways, to some extent.) Werbos and Dolmatova [2000] argued that quantum dynamics are not quite enough to justify quantum measurement here; one must also account for global boundary conditions, in order to derive quantum measurement, even as an approximation.

More recently, I have discussed these issues and related experiments with Yanhua Shih's group at Maryland (especially including Kim and Dolmatova-Werbos) and with Dowling and Strekalov at the Jet Propulsion Laboratory. There are new experiments in the pipeline exploiting the positional entanglement of light, originally inspired by theoretical suggestions from Klyshko. Klyshko created a partly-backwards-time picture of two-photon effects, which Shih has used to develop an extremely efficient engine for positional entanglement. That engine has been used, in turn, in experiments on quantum lithography and quantum imaging (among others) which go beyond anything achieved elsewhere. Preliminary discussions suggest that a more complete and more precise model of quantum measurement may be possible, based on a more careful analysis of the micro/macro interface. However, the details are beyond the scope of this paper, and further experiments are essential.

Our discussions so far have raised an interesting question: could it be possible to take the quantum-level backwards-time effects and raise them to macroscopic information flows? Years ago, many physicists felt that macroscopic "Schrodinger cat" experiments could not be performed in practice, for thermodynamic reasons. Yet just a few years ago, inspired by suggestions from Leggett and others, Nakamura proved that he actually could put a macroscopic object -- a SQUID device -- into a mixed state. I know of no reason as yet, in principle, why we could not do something similar for backwards-time causality. (Here I exclude apriori philosophical assumptions as a "reason.") Whether this is actually possible or not, it is important that we find out.

**2.3. Possible Implications for Computation**

Nothing in this paper should be interpreted as any kind of criticism of the excellent and exciting work now going on in Quantum Information Science (QIS), which includes quantum computing and related technologies. Leading spokesmen for that field, like Preskill of CalTech, have been very careful to keep their discussions and plans well grounded in what is known from solid experiments and well-grounded theoretical assumptions. Quantum computers have been made to work, albeit only on a small scale so far. Most of the existing work is based on entanglements of discrete variables like spin, as opposed to positional entanglement. Most of the recent work can be found simply by searching the web site xxx.lanl.gov.

This paper *does* suggest that quantum measurement effects are the key to the success of this work, relative to what would be possible with correlations across space alone. In practical QIS work, it is well known that the accuracy and control of quantum measurement (and setup) is a crucial limiting factor on what can be accomplished.

Many researchers have suggested, from time to time, that computation based on "classical" correlations (in electromagnetic fields, for example) could be used to achieve power similar to that of quantum computing. This paper suggests that they right in a sense -- but it also suggests that there is no essential physical difference between correlation-based optical computing and quantum computing. After all, they both use the same sort of light (albeit in different states). But the computational potential does depend on how well the various designs make use of quantum measurement effects and correlations over time. Paul Kwiat has analyzed the technical tradeoffs here in some detail, for the case of standard optical quantum computing, but the issues he has examined may have broader implications.

In the end, the backwards-time interpretation of quantum mechanics does suggest the possibility of new types of computing design beyond what mainstream physics has looked at so far.

Some enthusiasts for QIS, at high levels of the government, have argued that QIS might possibly give us a way to continue Moore's Law even after classical computing reaches its physical limits. But the bulk of QIS today depends on a handful of algorithms, important mainly to information security. (In the U.S., this is well-funded by the Department of Defense.) No one knows for sure whether quantum computing will ever be useful in a wide range of civilian, general-purpose computing applications. There has been important work in trying to develop "quantum simulation of quantum systems" (as first suggested by Feynmann), and there has been important work on quantum fast Fourier approaches; however, these are still niche markets.

A few authors, such as myself and Behrman et al [2002], have argued for years that someone might develop quantum neural networks (QNN) as a way to overcome this "algorithm bottleneck." Like conventional artificial neural networks (ANNs), these could be trained to approximate any nonlinear function and to support a wide variety of information-processing tasks. (For examples of such tasks, see the review paper posted at [www.iamcm.org.](www.iamcm.org.))

Recent research on ANNs has shown that conventional feedforward ANNs are actually very limited in the size of the problems they can handle. More general, more brain-like capabilities require new designs which, in turn, may require many iterations to settle down to a decision. (See Yang and Chua [1999] and [www.iamcm.org.](www.iamcm.org.)) Even with the human brain (which does not look like a quantum computing device) we are all familiar with the time it takes a human chess-player, for example, to settle down on a decision about what move to make next. But if hardware could be built in which a system achieves equilibrium by a kind of iteration between forward-time signals and backward-time signals, at the quantum level, a speed-up of such calculations might be possible. Even if the human brain is not actually a "quantum mind," we might be able to build devices which are. Nevertheless, the high density of today's VLSI suggests that it might be a long time before cellular QNNs are able to compete with the best classically based chip designs, like today's Cellular Neural Network. QNNs as such are straightforward in principle, but the full use of cross-time effects to accelerate QNNs would be far more difficult.

## 3. Review of Creation and Annihilation and Field Operators in QFT

Creation and annihilation operators are the main workhorses of QFT. They provide a compact formalism which is far more elegant and workable than traditional tools for working with Fock spaces or statistical moments. Their advantages are similar to the advantages of matrix algebra over older equation-by-equation representations of systems of linear equations.

To begin with, Fock space provides a more elegant way to represent wave functions like the ones illustrated in equations 13.

### 3.1. Definitions of creation and annihilation in the ODE case

Consider the example of an ODE system (i.e. d=0, so that $\underline{x}$ drops out of equations 11) in the case N=3. In this case, the symmetry of the function $\psi_n$ would imply, for example, that:

$$\psi_6(1, 1, 1, 2, 2, 3) = \psi_6(2, 1, 3, 2, 1, 1) \tag{15}$$

In fact, the symmetry implies that $\psi_6$ is really just a function of $N_1$, $N_2$ and $N_3$ -- the number of times that "1", "2" and "3" appear in the list of arguments. Thus a modern way to represent the wave function $\psi$ in this case is to use notation introduced by Dirac:

$$\psi = \sum_{N_1,N_2,N_3} c_{N_1,N_2,N_3} |N_1, N_2, N_3\rangle \tag{16}$$

where $c_{N1,N2,N3}$ is just the value of $\psi$ for the set of arguments $(N_1,N_2,N_3)$. The square norm of $\psi$ is then defined as:

$$|\psi|^2 = \sum_{N_1,N_2,N_3} |c_{N_1,N_2,N_3}|^2 \tag{17}$$

Notice that each possible combination of arguments is counted only *once* in the definition of the square norm! Also note that the square norm of the basis vector $|N_1 N_2 N_3\rangle$ is defined to equal 1. The basis vector $|000\rangle$ -- usually just written as $|0\rangle$ -- is called "the vacuum state." The norm of the vacuum state is also defined to equal 1; it should not be confused with the zero vector.

The creation operator $a_i^+$ is defined as the operator which increases $N_i$ by one and *also* multiplies the result by $N_i^{1/2}$ (using the new, larger value of $N_i$). Thus for example

$$a_2^+ |N_1, N_2, N_3\rangle = \sqrt{N_2+1} \, |N_1, N_2+1, N_3\rangle \tag{18}$$

In the simpler example where N=1, the states $|N_1\rangle$ form a simple ladder from $|0\rangle$ to $|1\rangle$ to $|2\rangle$ and so on; in that case, the creation operator $a_1^+$ is really just the (infinite) matrix:

$$\begin{bmatrix} ... & ... & ... & ... & ... \\ ... & 0 & \sqrt{3} & 0 & 0 \\ ... & 0 & 0 & \sqrt{2} & 0 \\ ... & 0 & 0 & 0 & 1 \\ ... & 0 & 0 & 0 & 0 \end{bmatrix} \tag{19}$$

The annihilation operator $a_i$ is just the Hermitian conjugate (transpose, in this case) of $a_i^+$.

It decreases $N_i$ by one and multiplies the result by $N_i^{1/2}$ (using the earlier, larger value of $N_i$). Thus for example:

$$a_2 |N_1, N_2, N_3\rangle = \sqrt{N_2} |N_1, N_2 - 1, N_3\rangle \tag{20}$$

Notice that when $N_2=0$, the vector on the right is just zero -- not the vacuum state, but zero. By combining equations 18 and 20 (and accounting for the changes in $N_2$) it is easy to see that:

$$a_2^+ a_2 |N_1, N_2, N_3\rangle = N_2 |N_1, N_2, N_3\rangle \tag{21}$$

Thus QFT defines a new operator, the number operator $N_i$, as:

$$N_i = a_i^+ a_i \tag{22}$$

Likewise, equations 18 and 20 imply:

$$a_2 a_2^+ |N_1, N_2, N_3\rangle = (N_2 + 1) |N_1, N_2, N_3\rangle \tag{23}$$

which implies, in general, that:

$$a_i a_i^+ = a_i^+ a_i + 1 \tag{24}$$

All of these examples and results extend in an obvious way to the case where N is any positive integer.

Based on these definitions, and on equation 16, we may even *define* the Fock-Hilbert space for $F^N(0)$ as the set of states:

(1) |0>

(2) all states generated from |0> by multiplying it (on the left, of course) by $a_1^+$, $a_2^+$,… or $a_N^+$ in some sequence;

(3) all linear combinations of the same which have finite norm

## 3.2. Fundamental basic tools

QFT makes very heavy use of the concept of *commutators*. The commutator of any two operators A and B is defined as:

$$[A, B] = AB - BA \qquad (25)$$

From this definition, we may easily derive the useful relations:

$$[AC, B] = A[C, B] + [A, B]C \qquad (26)$$

$$[A, BC] = B[A, C] + [A, B]C \qquad (27)$$

Equation 24 clearly tells us that:

$$[a_i, a_i^+] = 1 \qquad (28)$$

Going back to the definitions of $a_I$ and $a_i^+$, it is easy to see that the action of $a_i$ is independent of the actions of $a_j$ or of $a_j^+$, when i≠j. Thus it is easy to derive the well-known set of relations:

$$[a_i, a_j^+] = \delta_{ij} \qquad (29)$$

$$[a_i, a_j] = [a_i^+, a_j^+] = 0 \qquad (30)$$

where $\delta_{ij}$ is the "Kronecker delta." The Kronecker delta is defined to equal 1 when i=j, and zero in all other cases. When we confront any product of creation and annihilation operators, these equations allow us to derive an equivalent expression in which the creation operators are all to the left and the annihilation operators are all to the right. For example:

$$a_1^+ a_2^+ a_3 a_3^+ a_2 a_1 = a_1^+ a_2^+ \left([a_3, a_3^+] + a_3^+ a_3\right) a_2 a_1 = a_1^+ a_2^+ a_2 a_1 + a_1^+ a_2^+ a_3^+ a_3 a_2 a_1 \qquad (31)$$

When an operator is expressed as a sum of products, in which all the creation operators appear on the left-hand side of every product, we say that the operator has been expressed "in normal form."

It is also well-known from matrix algebra that:

$$(AB)^H = B^H A^H \tag{32}$$

(Our operators are not finite matrices; however, Kato[1995] has given a detailed mathematical justification for applying many of the tools of ordinary matrix theory to operators over Fock space.) Applying this to equation 25, we may deduce:

$$([A,B])^H = (AB - BA)^H = [B^H, A^H] = -[A^H, B^H] \tag{33}$$

QFT also makes very heavy use of the operator trace. The trace Tr(A) of an operator A is essentially just the sum of its diagonal elements. QFT makes very heavy use of the following identity from matrix algebra:

$$\text{Tr}(AB) = \text{Tr}(BA) \tag{34}$$

In the case where A or B is Hermitian (i.e., $A=A^H$ or $B=B^H$), Tr(AB) is really just the inner product of A and B when A and B are thought of as vectors in $N^2$-dimensional space. From matrix algebra, equation 34 holds even when A and B are not square matrices.

### 3.3. How commutators yield derivatives, using normal products

To begin with, consider the following lemma:

$$[a_j, (a_j^+)^m] = m(a_j^+)^{m-1} \tag{35}$$

for all nonnegative integers m. This is obvious for the case m=0, and it is identical to equation 24 in the case m=1. For m>1, it follows by mathematical induction, using equation 27 with the choice $B=a_j^+$ and $C=(a_j^+)^{m-1}$.

Next, for our purposes, we may define an analytic function $f(\underline{y})$ of the vector $\underline{y} \in C^N$ as any function which can be expressed as a convergent infinite sum:

$$f(\underline{y}) = \sum_{i_1,i_2,\ldots,i_N} C_{i_1,i_2,\ldots,i_N} y_1^{i_1} y_2^{i_2} \ldots y_N^{i_N} \tag{36}$$

Given such a function, we may define the operator version as:

$$f(\underline{a}^+) = \sum_{i_1,i_2,\ldots,i_N} C_{i_1,i_2,\ldots,i_N} (a_1^+)^{i_1} (a_2^+)^{i_2} \ldots (a_N^+)^{i_N} \tag{37}$$

Because $a_j$ commutes with all $a_k^+$ such that $k \neq j$, we may use our lemma to deduce:

$$[a_j, f(\underline{a}^+)] = \sum_{i_1,i_2,\ldots,i_N} C_{i_1,i_2,\ldots,i_N} (a_1^+)^{i_1} \ldots (a_{j-1}^+)^{i_{j-2}} [a_j, (a_j^+)^{i_j}](a_{j+1}^+)^{i_{j-1}} \ldots (a_N^+)^{i_N}$$
$$= \sum_{i_1,i_2,\ldots,i_N} i_j C_{i_1,i_2,\ldots,i_N} (a_1^+)^{i_1} \ldots (a_{j-1}^+)^{i_{j-2}} (a_j^+)^{i_j-1} (a_{j+1}^+)^{i_{j-1}} \ldots (a_N^+)^{i_N} \tag{38}$$

But notice that the right-hand side of equation 38 is just the operator version of the derivative of equation 36 with respect to $y_j$! Thus we may write:

$$[a_j, f(\underline{a}^+)] = \frac{\partial f}{\partial y_j}(\underline{a}^+) \tag{39}$$

The special case of this equation for N=1 appears in the well-known text on quantum optics by Walls and Milburn[1994]. The same approach (or a direct use of equation 33) yields:

$$[a_j^+, f(\underline{a})] = -\frac{\partial f}{\partial y_j}(\underline{a}) \tag{40}$$

Next consider the case of a function $f(\underline{y}, \underline{z})$ over $\underline{y}, \underline{z} \in C^N$ which is analytic in the sense that:

$$f(\underline{y},\underline{z}) = \sum_{\substack{i_1,i_2...i_N \\ j_1,j_2...j_N}} C_{j_1,j_2...j_N}^{i_1,i_2...i_N} y_1^{i_1}...y_N^{i_N} z_1^{j_1}...z_N^{j_N} \qquad (41)$$

For this function f, the corresponding function f($\underline{a}^+$,$\underline{a}$) *is not well-defined* in the general case. We certainly could (and will) define an operator by replacing each instance of $y_j$ in equation 41 with $a_j^+$, and replacing each instance of $z_j$ with $a_j$. But there are many other equally valid ways to represent the function f shown in equation 41; for example, the same function could be written with all the z's on the left and the y's on the right, because ordinary multiplication is commutative. If we then perform the same substitution into that other representation of f, we end up with an operator which does not equal the other operator. Knowing the function f($\underline{y}$, $\underline{z}$) by itself does not uniquely specify the corresponding operator.

To resolve these kinds of difficulties, physicists use a concept called *the normal product*. Any valid operator expression can be reduced to a sum of elementary products, simply by using the distributive law of multiplication, without using any knowledge about commutators. (An elementary product is just a string of creation and annihilation operators multiplied together.) And then, the normal product of an elementary product is the product of the same operators, rearranged so that the creation operators are all to the left of the annihilation operators. For example:

$$N(a_1 a_2^+ a_1^+) = \; : a_1 a_2^+ a_1^+ : \; = a_1^+ a_2^+ a_1 \qquad (42)$$

Notice that the normal product of $a_1 a_2^+ a_1^+$ does not equal $a_1 a_2^+ a_1^+$. Also note that we don't care what the order of operators is amongst the creation operators themselves, because they all commute with each other, and likewise amongst the annihilation operators. The "N" notation and the "::" notation for normal products are both commonly used in physics.

The expression f($\underline{a}^+$, $\underline{a}$) is not well-defined, for the reasons just given. However, the normal product *is* well-defined, and I will denote it as:

$$:f(\underline{a}^+,\underline{a}): = f_n(\underline{a}^+,\underline{a}) = \sum_{\substack{i_1,i_2...i_N \\ j_1,j_2...j_N}} C^{i_1,i_2...i_N}_{j_1,j_2...j_N} (a_1^+)^{i_1}...(a_N^+)^{i_N} a_1^{j_1}...a_N^{j_N} \qquad (43)$$

Because $a_j$ commutes with all of the annihilation operators here, we obtain the same sort of result as in equation 38:

$$[a_j, f_n(\underline{a}^+,\underline{a})] = \sum_{\substack{i_1,i_2...i_N \\ j_1,j_2...j_N}} i_j C^{i_1,i_2...i_N}_{j_1,j_2...j_N} (a_1^+)^{i_1}...(a_{j-1}^+)^{i_{j-1}} (a_j^+)^{i_j-1} (a_{j+1}^+)^{i_{j+1}}...(a_N^+)^{i_N} a_1^{j_1}...a_N^{j_N} \qquad (44)$$

But this is just the normal operator version of the derivative of equation 41 with respect to $y_j$! Thus:

$$[a_j, f_n(\underline{a}^+,\underline{a})] = \left(\frac{\partial f}{\partial y_j}\right)_n (\underline{a}^+,\underline{a}) \qquad (45)$$

Likewise:

$$[a_j^+, f_n(\underline{a}^+,\underline{a})] = -\left(\frac{\partial f}{\partial z_j}\right)_n (\underline{a}^+,\underline{a}) \qquad (46)$$

### 3.4. Field operators $\Phi_j$ and $\Pi_j$ and differentiation

Bosonic QFT is constructed from fundamental operators $\Phi_j(\underline{x})$ and $\Pi_j(\underline{x})$ which take on a simple form in the ODE case:

$$\Phi_j = \frac{1}{\sqrt{2}}(a_j + a_j^+) \qquad (47)$$

$$\Pi_j = \frac{1}{i\sqrt{2}}(a_j - a_j^+) \tag{48}$$

If we insert these definitions and exploit equations 29 and 30, we may easily derive the standard relations:

$$[\Phi_j, \Pi_k] = \delta_{jk}(\frac{1}{\sqrt{2}})(\frac{1}{i\sqrt{2}})(-2) = i\delta_{jk} \tag{49}$$

$$[\Phi_j, \Phi_k] = [\Pi_j, \Pi_k] = 0 \tag{50}$$

Using the same argument by induction we used to prove equation 35, we may deduce:

$$[\Phi_j, (\Pi_k)^m] = i\delta_{jk} m (\Pi_k)^{m-1} \tag{51}$$

Now let us address the problem of how to compute derivatives for the operator version of an analytic function $g(\underline{\varphi}, \underline{\pi})$, where $\underline{\varphi}, \underline{\pi} \in R^N$. Substituting in from equations 47 and 48, and recalling the discussion of normal products in section 3.3, we may write:

$$g_n(\underline{\Phi}, \underline{\Pi}) = : g\left(\frac{1}{\sqrt{2}}(\underline{a} + \underline{a}^+), \frac{1}{i\sqrt{2}}(\underline{a} - \underline{a}^+)\right): \tag{52}$$

Let us define a new function f as:

$$f(\underline{y}, \underline{z}) = g\left(\frac{1}{\sqrt{2}}(\underline{z} + \underline{y}), \frac{1}{i\sqrt{2}}(\underline{z} - \underline{y})\right) \tag{53}$$

Note that $f_n(\underline{a}^+, \underline{a})$ and $g_n(\underline{\Phi}, \underline{\Pi})$ are exactly the same operators, when expressed as functions of the elementary creation and annihilation operators.

Applying equation 45 and the chain rule for differentiation, we may deduce:

$$[a_j, g_n(\underline{\Phi}, \underline{\Pi})] = \left(\frac{\partial f}{\partial y_j}\right)_n (\underline{a}^+, \underline{a}) = \frac{1}{\sqrt{2}}\left(\frac{\partial g}{\partial \varphi_j}\right)_n (\underline{\Phi}, \underline{\Pi}) - \frac{1}{i\sqrt{2}}\left(\frac{\partial g}{\partial \pi_j}\right)_n (\underline{\Phi}, \underline{\Pi}) \tag{54}$$

Likewise, applying equation 46, we may deduce:

$$[a_j^+, g_n(\underline{\Phi}, \underline{\Pi})] = -\left(\frac{\partial f}{\partial z_j}\right)_n (\underline{a}^+, \underline{a}) = -\frac{1}{\sqrt{2}}\left(\frac{\partial g}{\partial \varphi_j}\right)_n (\underline{\Phi}, \underline{\Pi}) - \frac{1}{i\sqrt{2}}\left(\frac{\partial g}{\partial \pi_j}\right)_n (\underline{\Phi}, \underline{\Pi}) \qquad (55)$$

Adding equations 54 and 55 together, and dividing by sqrt(2), we deduce:

$$[\Phi_j, g_n(\underline{\Phi}, \underline{\Pi})] = -\frac{1}{\sqrt{2}}(2)\frac{1}{i\sqrt{2}}\left(\frac{\partial g}{\partial \pi_j}\right)_n (\underline{\Phi}, \underline{\Pi}) = i\left(\frac{\partial g}{\partial \pi_j}\right)_n (\underline{\Phi}, \underline{\Pi}) \qquad (56)$$

Subtracting equation 55 from equation 54, and dividing by i*sqrt(2), we may likewise deduce:

$$[\Pi_j, g_n(\underline{\Phi}, \underline{\Pi})] = -i\left(\frac{\partial g}{\partial \varphi_j}\right)_n (\underline{\Phi}, \underline{\Pi}) \qquad (57)$$

Equations 56 and 57 will be very important to section 4.

**3.5. Hamiltonian operators and zero-point energy in bosonic QFT**

To complete the specification of bosonic QFT in the ODE case, we need only define the operator $H_n$ in equations 1 and 2.

Normally, in texts on QFT, the Liouville equation is written with "H" instead of "$H_n$". H is generally described as the "quantized" version of the energy function H of equation 6. In other words, if H is the analytic function in equation 6, the operator used in the Liouville equation is said to be H($\underline{\Phi}$, $\underline{\Pi}$), *which is not well-defined in the general case*. However, if one examines the subsequent calculations of Feynmann integrals, or even just the discussions of interaction Hamiltonians, it turns out that the normal product Hamiltonian, $H_n(\underline{\Phi}, \underline{\Pi})$ is the operator actually used. The difference between the operator H initially used and the operator $H_n$ is declared to be unobservable and is thrown out in early stages of the discussion. Textbooks vary a great deal in how they discuss the rationale for actually using $H_n$ in the real calculations.

Unfortunately, the unobserved, cosmetic terms $H-H_n$ have caused great excitement among those looking for new sources of energy or warp drives or the like -- worthy goals which these particular terms cannot help with. Some authors have claimed that these "vacuum energy" or "zero point" terms are necessary to explain the Lamb shift effect in atomic spectra; however, the "vacuum fluctuation" Feynmann diagrams used in explaining the Lamb shift are based on the normal form Hamiltonian. Others have argued that the famous Casimir effect can only be explained by invoking zero point terms; however, Landau showed long ago that the observed Casimir effects are precisely equal to what one would predict from ordinary dipole-dipole interactions. [Hoye et al, 2001]. Finally, some mid-level books and papers on quantum optics try to use semi-classical models to explain phenomena like spontaneous emission and noise suppression in lasers; they are forced to assume zero-point effects to make this work. However, in full-fledged quantum optics, based on the normal form Hamiltonian, there is no need to use such tricks.

**3.6. Extension to the PDE case**

Weinberg [1995] gives the PDE version of many of the relations discussed above. For example, in chapter 4, he states (in different notation):

$$[a_i(\underline{x}), a_j^+(\underline{y})] = \delta_{ij}\delta^d(\underline{x}-\underline{y}), \tag{58}$$

where the rightmost δ is the standard "Dirac delta function," a distribution whose integral with respect to $\underline{x}$ equals 1 over any interval which includes $\underline{y}$; its integral equals zero over any interval which does not include $\underline{y}$. (I have inserted the superscript "d" to make this valid for the case of d-dimensional space.) In equations 7.2.29 and 7.2.30, he defines the field operators $\Phi_j(\underline{x})$ and $\underline{\Pi}_j(\underline{x})$; in my notation, generalizing from 3 dimensions of space to d dimensions of space, Weinberg's definitions become:

$$\Phi_j(\underline{x}) = \Phi_j^+(\underline{x}) + \Phi_j^-(\underline{x}) \tag{59}$$

$$\Pi_j(\underline{x}) = \Pi_j^+(\underline{x}) + \Pi_j^-(\underline{x}) \tag{60}$$

where:

$$\Phi_j^-(\underline{x}) = \left(\Phi_j^+(\underline{x})\right)^H \tag{61}$$

$$\Pi_j^-(\underline{x}) = \left(\Pi_j^+(\underline{x})\right)^H \tag{62}$$

$$\Phi_j^+(\underline{x}) = c \int \frac{e^{i\underline{p}\cdot\underline{x}} a_j(\underline{p})}{\sqrt{w_j(\underline{p})}} d^d \underline{p} \tag{63}$$

$$\Pi_j^+(\underline{x}) = -ic \int \left(\sqrt{w_j(\underline{p})}\right) e^{i\underline{p}\cdot\underline{x}} a_j(\underline{p}) d^d \underline{p} \tag{64}$$

$$c = \frac{1}{\sqrt{2(2\pi)^d}} \tag{65}$$

Weinberg actually provides several different varieties of field operators, in order to be explicit about the various possibilities for how the fields change under a Lorentz transformation; however, as discussed in section 2.1, we do not need to be so explicit here.

Finally, in his equations 7.2.31-33, he provides the obvious generalizations of equations 49 and 50. To within certain technicalities, these generalizations are:

$$[\Phi_i(\underline{x}), \Pi_j(\underline{y})] = i\delta_{ij}\delta^d(\underline{x}-\underline{y}) \tag{66}$$

$$[\Phi_i(\underline{x}), \Phi_j(\underline{y})] = [\Pi_i(\underline{x}), \Pi_j(\underline{y})] = 0 \tag{67}$$

Equations 58 and 66 make it clear that the continuous arguments of $\Phi$ and $\Pi$ operate in the same basic way as the integer or spin arguments do. Thus we would expect all of the commutator results of this section to go through in the PDE case; however, to be truly rigorous, we would need to invoke the formal theory of distributions. Intuitively, it seems clear that the usual process of defining fields over a discrete lattice and taking the limit as lattice spacing goes to zero should work here, as it generally does in ordinary QFT.

## 4. Derivation of Equations 6, 8 and 9 (ODE/QFT Equivalence)

### 4.1. Definition of the classical density matrix ρ

The classical density matrix ρ and its properties are the most important new concepts in this paper. In the ODE case, ρ(t) is a way of encoding all of the statistical correlations of the vectors $\underline{\varphi}(t)$ and $\underline{\pi}(t)$ describing the state of a classical system at time t.

Efforts to understand the statistics of classical fields have a long history which are beyond the scope of this paper; however, some historical concepts are essential to understand the new work.

For centuries, statisticians have studied how probability distributions Pr(φ) can be represented and studied in terms of statistical moments, correlations or cumulants. For example, the set of real numbers

$$u_{i_1,i_2,...,i_N} = \left\langle \varphi_1^{i_1} \varphi_2^{i_2} ... \varphi_N^{i_N} \right\rangle \tag{68}$$

includes the statistical mean and (implicitly) the covariance, skewness and kurtosis, etc. The set of all such u values for a given probability distribution may be seen, mathematically, as a kind of a real wave function, as a function over the Fock space $F^N(d)$ discussed in sections 2.1! Unfortunately, the moments do not form a vector in the Fock-Hilbert space discussed in section 3, because they do not have a bounded norm.

Many classical physicists have tried to "close turbulence" by deriving linear or nonlinear dynamical equations for the statistical moments. However, when the statistical moments u are used directly – without scaling and without exploiting their special properties – the dynamical equations often display a problem of infinite regress, explicitly or implicitly. Many of those efforts are analogous to older textbooks which try to solve simultaneous linear equations by writing each equation separately and manipulating them one by one; by shifting over to Fock space concepts from QFT, we can achieve a simplicity and power analogous to the power which matrix algebra provides in working with systems of linear equations.

In order to represent the moments as a vector in Fock-Hilbert space, Werbos [1993] and Werbos and Dolmatova [2000] suggested that we scale them:

$$v_{i_1,i_2,...,i_n} = \frac{u_{i_1,i_2,...,i_n}}{\sqrt{i_1! i_2! ... i_n!}}, \tag{69}$$

which is equivalent to the more compact and modern definition:

$$\underline{v}(\underline{\varphi}) = \exp\left(\sum_{j=1}^{n} \varphi_j a_j^+\right) | 0 > \tag{70}$$

This lets us represent a probability distribution Pr($\underline{\varphi}$) by:

$$\underline{v} = <\underline{v}(\underline{\varphi})> = \int \underline{v}(\underline{\varphi}) \Pr(\underline{\varphi}) d^n \varphi \tag{71}$$

Note that equations 69 and 70 actually make sense even for vectors $\underline{\varphi} \in C^N$.

To avoid confusion, I will now consider the case where $\underline{y}$ is used to represent such a complex vector. For any complex vector $\underline{y}$, we may easily compute the square norm of $\underline{v}(\underline{y})$ as defined by equation 17:

$$\left|\underline{v}(\underline{y})\right|^2 = \sum_{i_1,i_2,\ldots,i_N} \left|\frac{y_1^{i_1} y_2^{i_2} \ldots y_N^{i_N}}{\sqrt{i_1! i_2! \ldots i_N!}}\right|^2 = \sum_{i_1,i_2,\ldots,i_N} \frac{(|y_1|^2)^{i_1}}{i_1!} \cdot \frac{(|y_2|^2)^{i_2}}{i_2!} \cdots \frac{(|y_N|^2)^{i_N}}{i_N!}$$
$$= \exp(|y_1|^2) \cdot \exp(|y_2|^2) \cdots \exp(|y_N|^2) = \exp(|\underline{y}|^2) \tag{72}$$

Thus $\underline{v}(\underline{y})$ will always have a finite norm. However, the norm is still not equal to 1 (as it is for wave functions used in QFT). More seriously, equation 71 only represents the statistics of $\underline{\varphi}$, not the statistics of the entire state ($\underline{\varphi}$, $\underline{\pi}$)! Thus to represent the statistical moments of the state of the system shown in equation 7, I propose that we start from the vector:

$$\underline{w}(\underline{\varphi},\underline{\pi}) = \frac{\underline{v}(\underline{\varphi}+i\underline{\pi})}{|\underline{v}(\underline{\varphi}+i\underline{\pi})|} = \exp\left(\frac{1}{\sqrt{2}} \sum_{j=1}^{n} \left((\varphi_j + i\pi_j) a_j^+ - \tfrac{1}{2\sqrt{2}}(\varphi_j^2 + \pi_j^2)\right)\right)|0\rangle \tag{73}$$

The classical density matrix is then defined by

$$\rho = \iint \underline{w}(\underline{\varphi},\underline{\pi}) \underline{w}^H(\underline{\varphi},\underline{\pi}) \Pr(\underline{\varphi},\underline{\pi}) d^n \varphi \, d^n \pi \quad , \tag{74}$$

where the superscript H denotes the usual Hermitian conjugate. Vectors $\underline{w}$ which represent a specific state ($\underline{\varphi},\underline{\pi}$) as in equation 73 will be called "pure states." Matrices $\rho$ which represent an ensemble of states as in equation 74 will be called "physically realizable." Equation 74 looks like the usual definition of the Glauber-Sudarshan representation described by Walls and Milburn[1994], but here the probabilities are real probabilities and we are using them to define a mapping from classical field ensembles to a new object, ρ, over Fock space.

The vector **w**(**φ**, **π**) in equation 73 is an extension of the earlier concept of **w**(**φ**) used in Werbos [1993] and Werbos and Dolmatova [2000]. It is also related to the description of coherent states in Walls and Milburn [1994]. In fact, **w**(**φ**, **π**) can be defined in an equivalent but alternative way by using unitary operators very similar to those of Walls and Milburn:

$$\underline{w}(\underline{\varphi},\underline{\pi}) = \prod D_j(\varphi_j + i\pi_j) | 0 >= \exp\left(\frac{1}{\sqrt{2}} \sum_{j=1}^{n} \left((\varphi_j + i\pi_j)a_j^+ - (\varphi_j - i\pi_j)a_j\right)\right)|0\rangle \quad (75)$$

(Note that physics has adopted the convention of defining exp(A) as the ordinary Taylor series expansion in A.) These close connections suggest that the "pure states" of the classical theory do map into the "coherent states" of quantum optics, with a few caveats related to quantum measurement, beyond the scope of this paper.

**4.2. "Measurement" properties and of ρ and <u>w</u>(φ, π) (equations 8 and 9)**

First let me prove the following lemma for all vectors **φ** and **π** ∈ $R^N$:

$$a_j \underline{w}(\underline{\varphi},\underline{\pi}) = \left(\frac{\varphi_j + i\pi_j}{\sqrt{2}}\right) \underline{w}(\underline{\varphi},\underline{\pi}) \quad (76)$$

To prove this without making the equations too complicated, I will first define

the intermediate quantity:

$$W(\underline{\varphi},\underline{\pi}) = \frac{1}{\sqrt{2}} \sum_{j=1}^{n} \left((\varphi_j + i\pi_j)a_j^+ - \tfrac{1}{2}(\varphi_j^2 + \pi_j^2)\right) \tag{77}$$

Then, from equation 73 and from the definition of a commutator, we get:

$$a_j \underline{w}(\underline{\varphi},\underline{\pi}) = a_j \exp(W(\underline{\varphi},\underline{\pi}))|0> = [a_j, \exp(W(\underline{\varphi},\underline{\pi}))]|0> + \exp(W(\underline{\varphi},\underline{\pi}))a_j|0> \tag{78}$$

The rightmost term here is zero, because $a_j|0>=0$. The other term may be evaluated by using equation 45 (and implicitly associating each $a_k^+$ with a "$y_k$"), which yields:

$$a_j \underline{w}(\underline{\varphi},\underline{\pi}) = (\frac{\partial}{\partial a_j^+} \exp(W(\underline{\varphi},\underline{\pi})))|0> = \frac{\partial W(\underline{\varphi},\underline{\pi})}{\partial a_j^+} \exp(W(\underline{\varphi},\underline{\pi})))|0>$$
$$= \frac{\partial W(\underline{\varphi},\underline{\pi})}{\partial a_j^+} \underline{w}(\underline{\varphi},\underline{\pi}) \tag{79}$$

This reduces to equation 76, when we simply differentiate equation 77 to calculate the required derivative. Q.E.D. Note that our use of equation 45 is legitimate here because W and $\underline{w}$ are all in normal form; they are made up entirely of creation operators.

Next, from equation 76, equation 34, and the definition of $\rho$ we may deduce, for any statistical ensemble of states:

$$Tr(\rho \Phi_i) = \frac{1}{\sqrt{2}} Tr(\rho a_i + \rho a_i^+) = \frac{1}{\sqrt{2}} <\underline{w}^H a_i \underline{w}> + \frac{1}{\sqrt{2}} <(a_i \underline{w})^H \underline{w}>$$
$$= \frac{1}{\sqrt{2}} <\frac{\varphi_i + i\pi}{\sqrt{2}}>_i + \frac{1}{\sqrt{2}} <\frac{\varphi_i - i\pi}{\sqrt{2}}>_i = <\varphi_i>. \tag{80}$$

and similarly for $\Pi_i$ and $\pi_i$. *Note that we have just proven equation 8.*

To conclude this section, I will now prove equation 9. For any statistical ensembles of states and any analytic function g, I claim that:

$$Tr(\rho g_n(\underline{\Phi},\underline{\Pi})) = <g(\underline{\varphi},\underline{\pi})> \tag{81}$$

To prove this, I will first prove that it holds for any function g of the following form:

$$g(\underline{\varphi},\underline{\pi}) = \left(\frac{\varphi_1 - i\pi_1}{\sqrt{2}}\right)^{i_1} \left(\frac{\varphi_2 - i\pi_2}{\sqrt{2}}\right)^{i_2} \cdots \left(\frac{\varphi_N - i\pi_N}{\sqrt{2}}\right)^{i_N} \left(\frac{\varphi_1 + i\pi_1}{\sqrt{2}}\right)^{j_1} \cdots \left(\frac{\varphi_N + i\pi_N}{\sqrt{2}}\right)^{j_N} \quad (82)$$

I begin with this special case because it is easy to compute the normal product for this case. If we substitute in from the definitions of $\Phi_j$ and $\Pi_j$ in equations 47 and 48, we see that:

$$g_n(\underline{\Phi},\underline{\Pi}) = N\left(\left(\tfrac{1}{\sqrt{2}}(\Phi_1 - i\Pi_1)\right)^{i_1} \cdots \left(\tfrac{1}{\sqrt{2}}(\Phi_N - i\Pi_N)\right)^{i_N} \left(\tfrac{1}{\sqrt{2}}(\Phi_1 + i\Pi_1)\right)^{j_1} \cdots \left(\tfrac{1}{\sqrt{2}}(\Phi_N + i\Pi_N)\right)^{j_N}\right)$$
$$= N\left((a_1^+)^{i_1} \cdots (a_N^+)^{i_N} a_1^{j_1} \cdots a_N^{j_N}\right) = (a_1^+)^{i_1} \cdots (a_N^+)^{i_N} a_1^{j_1} \cdots a_N^{j_N} \quad (83)$$

For such a function g, we can use equations 34 and 74 to derive:

$$\text{Tr}(\rho g_n(\underline{\Phi},\underline{\Pi})) = \text{Tr}(\rho (a_1^+)^{i_1} \cdots (a_N^+)^{i_N} a_1^{j_1} \cdots a_N^{j_N})$$
$$= \text{Tr}(a_1^{j_1} \cdots a_N^{j_N} \rho (a_1^+)^{i_1} \cdots (a_N^+)^{i_N}) \quad (84)$$
$$= \int \text{Tr}(a_1^{j_1} \cdots a_N^{j_N} \underline{w}(\underline{\varphi},\underline{\pi}) \underline{w}^H(\underline{\varphi},\underline{\pi}) (a_1^+)^{i_1} \cdots (a_N^+)^{i_N}) \Pr(\underline{\varphi},\underline{\pi}) d^N\underline{\varphi}\, d^N\underline{\pi}$$

From equation 76 and its Hermitian conjugate, this can be further reduced to:

$$\int \text{Tr}\left(\left(\tfrac{\varphi_1 + i\pi_1}{\sqrt{2}}\right)^{j_1} \cdots \left(\tfrac{\varphi_N + i\pi_N}{\sqrt{2}}\right)^{j_N} \underline{w}\,\underline{w}^H \left(\tfrac{\varphi_1 - i\pi_1}{\sqrt{2}}\right)^{i_1} \cdots \left(\tfrac{\varphi_N - i\pi_N}{\sqrt{2}}\right)^{i_N}\right) \Pr(\underline{\varphi},\underline{\pi}) d^N\underline{\varphi}\, d^N\underline{\pi}$$

$$\int \text{Tr}(\underline{w}\,\underline{w}^H) \left(\tfrac{\varphi_1 + i\pi_1}{\sqrt{2}}\right)^{j_1} \cdots \left(\tfrac{\varphi_N + i\pi_N}{\sqrt{2}}\right)^{j_N} \left(\tfrac{\varphi_1 - i\pi_1}{\sqrt{2}}\right)^{i_1} \cdots \left(\tfrac{\varphi_N - i\pi_N}{\sqrt{2}}\right)^{i_N} \Pr(\underline{\varphi},\underline{\pi}) d^N\underline{\varphi}\, d^N\underline{\pi}$$

$$= \left\langle \left(\tfrac{\varphi_1 + i\pi_1}{\sqrt{2}}\right)^{j_1} \cdots \left(\tfrac{\varphi_N + i\pi_N}{\sqrt{2}}\right)^{j_N} \left(\tfrac{\varphi_1 - i\pi_1}{\sqrt{2}}\right)^{i_1} \cdots \left(\tfrac{\varphi_N - i\pi_N}{\sqrt{2}}\right)^{i_N} \right\rangle = <g(\underline{\varphi},\underline{\pi})> \quad (85)$$

Q.E.D. In equation 85, I wrote "$\underline{w}$" instead of $\underline{w}(\underline{\varphi},\underline{\pi})$ simply to hold down the size of the equation. In the last line, I exploited the fact that $\text{Tr}(\underline{w}\underline{w}^H)=\text{Tr}(\underline{w}^H\underline{w})=|\underline{w}|^2=1$, as well as the fact that $\text{Tr}(cA)=c\text{Tr}(A)$ for any scalar c.

The general case of the theorem (for all analytic functions g) falls out from linearity.

First, we need to see that any analytic function g can be expressed as a linear combination (possibly infinite) of functions $g_\alpha$ of the form of equation 82. There are several ways to show this. For example, any analytic function g can be expressed as a linear combination of terms made up of products of $\varphi_j$ and $\pi_j$ for various j. In each product, each term $\varphi_j$ can be re-expressed as $(1/2)(\varphi_j+i\pi_j)+(1/2)(\varphi_j-i\pi_j)$, and each term $\pi_k$ can be re-expressed as $(1/2i)(\varphi_j+i\pi_j)-(1/2i)(\varphi_j-i\pi_j)$. With that substitution, and distribution of multiplication, each product term is expressed as a linear combination of product terms of the form of equation 82. Since a linear combination of a linear combination is still a linear combination, g itself is therefore expressible as a linear combination of components $g_\alpha$ of the form shown in equation 82.

When g is expressed as such a sum, our lemma lets us deduce that:

$$\mathrm{Tr}(\rho g_n(\Phi,\Pi)) = \mathrm{Tr}\left(\rho\left(\sum_\alpha C_\alpha g_{\alpha n}(\Phi,\Pi)\right)\right) = \sum_\alpha C_\alpha \mathrm{Tr}(\rho g_{\alpha n}(\Phi,\Pi))$$
$$= \sum_\alpha C_\alpha <g_\alpha(\underline{\varphi},\underline{\pi})> = \left\langle \sum_\alpha C_\alpha g_\alpha(\underline{\varphi},\underline{\pi}) \right\rangle = <g(\underline{\varphi},\underline{\pi})> \quad (86)$$

Q.E.D. This approach can also be used to generate an alternative (lengthier) proof of equations 56 and 57.

### 4.3. Dynamics of ρ (proof of equation 6)

### 4.3.1. Preliminary results for $\Phi_j$, $\Pi_j$, $a_j$ and $a_j^+$

The dynamical equations for the classical system in equation 7 are well-known, and can be derived directly as the Lagrange-Euler equations for equation 7:

$$\dot{\varphi}_j = \frac{\partial H}{\partial \pi_j} \tag{87}$$

$$\dot{\pi}_j = -\frac{\partial H}{\partial \varphi_j} \tag{88}$$

Taking the expectation value of equation 87, and exploiting equations 56 and 81, we may deduce:

$$<\dot{\varphi}_j> = <\frac{\partial H}{\partial \pi_j}> = \text{Tr}\left(\rho \left(\frac{\partial H}{\partial \pi_j}\right)_n (\Phi, \Pi)\right) = -i\,\text{Tr}(\rho[\Phi_j, H_n(\Phi, \Pi)]) \tag{89}$$

Likewise:

$$<\dot{\pi}_j> = -i\,\text{Tr}(\rho[\Pi_j, H_n(\Phi, \Pi)]) \tag{90}$$

Following the approach of equations 52 and 53, we may define new variables:

$$z_j = \tfrac{1}{\sqrt{2}}(\varphi_j + i\pi_j) \tag{91}$$

$$y_j = \tfrac{1}{\sqrt{2}}(\varphi_j - i\pi_j) \tag{92}$$

If we multiply equation 90 by i, add the result to equation 89, and divide by sqrt(2), we may derive:

$$<\dot{z}_j> = -i\,\text{Tr}\left(\rho\left[\frac{\Phi_j + i\Pi_j}{\sqrt{2}}, H_n(\Phi, \Pi)\right]\right) \tag{93}$$

Using the definitions of $\Phi_j$ and $\Pi_j$ (equations 47 and 48), we may calculate:

$$\frac{\Phi_j + i\Pi_j}{\sqrt{2}} = \tfrac{1}{2}((a_j + a_j^+) + (a_j - a_j^+)) = a_j \tag{94}$$

Thus equation 94 reduces to:

$$< \dot{z}_j > = -i \operatorname{Tr}\left(\rho[a_j, H_n(\underline{\Phi}, \underline{\Pi})]\right) \tag{95}$$

In the same way, we may deduce:

$$< \dot{y}_j > = -i \operatorname{Tr}\left(\rho[a_j^+, H_n(\underline{\Phi}, \underline{\Pi})]\right) \tag{96}$$

**4.3.2. Proof of equation 6 for statistics of ODE**

From the definitions given in equations 1 and 5, we may deduce:

$$\left.\frac{d}{dt}\right|_{t=0} Q_j(t) = i[\Phi_j, H_n] \tag{97}$$

From equation 8, we may deduce that:

$$< \dot{\varphi}_j > = \frac{d}{dt} \operatorname{Tr}(\rho(t)\Phi_j) \tag{98}$$

If we substitute these equations into equation 6, we see that equation 6 is simply the same as equation 89 – which we have already proven – for the case t=0. The same logic works for $P_j$ and $\Pi_j$ as well. Q.E.D.

**V. Extension to d+1-Dimensional PDE Case**

This section will discuss a few key details of how to extend the results of the previous section to the d+1 dimensional case. (See Werbos and Dolmatova [2000] for some alternative statistical representations which may also be useful in the general case.) First let us try to analyze the statistical dynamics of the following class of CFT, expressed in Hamiltonian form:

$$L = \tfrac{1}{2}\sum_{j=1}^{n}\left(\dot{\varphi}_j \pi_j - \dot{\pi}_j \varphi_j - |\nabla \varphi_j|^2 - m_j^2 \varphi_j^2\right) - f(\underline{\varphi}, \underline{\pi}, \nabla \underline{\varphi}), \tag{99}$$

where $\underline{\varphi} \in R^n$ and f is an analytic function. In theoretical physics, we usually restrict our attention to the special case where the mathematical vector $\underline{\varphi}$ is actually composed of relativistic scalars, vectors, isovectors, tensors, etc.; however it is easier here to address the general case. Whenever the original Lagrangian is relativistically invariant, then the equivalent QFT will also yield predictions consistent with relativity.

Note that equation 99 has some redundancy in it. Any dynamical system with a Lagrangian of this form can be expressed in many equivalent ways; for example, we can change the "mass" terms $m_j$, and balance those changes by changes in f, to arrive at a "different" Lagrangian which is actually the same functional of the fields. (In principle, we could even make the mass terms functions of $|\underline{p}|$, after a Fourier transform; however, I have no need of that here.) There is no one "right" choice between these alternatives; all are valid, so long as they are consistent with later stages of analysis.

CFT then gives us the classical Hamiltonian density and Lagrange-Euler equations:

$$H = \tfrac{1}{2}\sum_{j=1}^{n}\left(|\nabla \varphi_j|^2 + m_j^2 \varphi_j^2\right) + f(\underline{\varphi}, \underline{\pi}, \nabla \underline{\varphi}) \tag{100}$$

$$\dot{\varphi}_j = \frac{\partial f}{\partial \pi_j} \tag{101}$$

$$\dot{\pi}_j = \Delta \varphi_j - m_j^2 \varphi_j - \frac{\partial f}{\partial \varphi_j} - \left(\nabla \cdot \frac{\delta f}{\delta(\nabla \varphi_j)}\right) \tag{102}$$

Let S(t) denote the state of this dynamical system at any time t; S consists of all of the values of $\underline{\varphi}(\underline{x},t)$ and $\underline{\pi}(\underline{x},t)$ across all points in space $\underline{x}$ at time t. Then

the classical density matrix may be defined by the functional integral:

$$\rho = \int \frac{\underline{v}(S)\underline{v}^H(S)}{|\underline{v}(S)|^2} \Pr(S) d^\infty S, \tag{103}$$

where:

$$\underline{v}(S) = \exp\left( c \sum_{j=1}^{n} \int (\theta_j(\underline{p}) + i\tau_j(\underline{p})) a_j^+(\underline{p}) d^d \underline{p} \right) |0\rangle, \tag{104}$$

where d is the number of spatial dimensions (i.e. $\underline{x} \in R^d$) and:

$$\theta_j(\underline{p}) = \sqrt{w_j(\underline{p})} \int e^{-i\underline{p}\cdot\underline{y}} \varphi_j(\underline{y}) d^d \underline{y} \tag{105}$$

$$\tau_j(\underline{p}) = \frac{1}{\sqrt{w_j(\underline{p})}} \int e^{-i\underline{p}\cdot\underline{y}} \pi_j(\underline{y}) d^d \underline{y} \tag{106}$$

$$w_j(\underline{p}) = \sqrt{m_j^2 + |\underline{p}|^2} \tag{107}$$

It is straightforward but tedious to then prove (for d=3) that:

$$\text{Tr}(\rho \Phi_j(\underline{x})) = <\varphi_j(\underline{x})> \tag{108}$$

$$\text{Tr}(\rho \Pi_j(\underline{x})) = <\pi_j(\underline{x})> \tag{109}$$

where the field operators $\Phi_j$ and $\Pi_j$ are defined *exactly* as in Weinberg [1995], equations 7.2.29 and 7.2.30, which in my notation may be written:

$$\Phi_j(\underline{x}) = \Phi_j^+(\underline{x}) + \Phi_j^-(\underline{x}) \tag{110}$$

$$\Pi_j(\underline{x}) = \Pi_j^+(\underline{x}) + \Pi_j^-(\underline{x}) \tag{111}$$

where:

$$\Phi_j^-(\underline{x}) = \left(\Phi_j^+(\underline{x})\right)^H \tag{112}$$

$$\Pi_j^-(\underline{x}) = \left(\Pi_j^+(\underline{x})\right)^H \tag{113}$$

$$\Phi_j^+(\underline{x}) = c \int \frac{e^{i\underline{p}\cdot\underline{x}} a_j(\underline{p})}{\sqrt{w_j(\underline{p})}} d^d \underline{p} \qquad (114)$$

$$\Pi_j^+(\underline{x}) = -ic \int \left(\sqrt{w_j(\underline{p})}\right) e^{i\underline{p}\cdot\underline{x}} a_j(\underline{p}) d^d \underline{p} \qquad (115)$$

In actuality, to perform these calculations, I first treated c as an unknown parameter, and solved for $\Phi_j^+(\underline{x})\underline{v}(S)=(\varphi_j(\underline{x})+ \text{imaginary})\underline{v}(S)$, and likewise for $\Pi$. (I found it convenient as well to replace "p" by "q" in equations 104-107, before proceeding. Those definitions could be expressed a bit more elegantly, but the form given here was convenient for derivation.)

With these definitions, one would expect the logic behind equation 9 to go through as before, yielding the exact same equation for the PDE case! The standard commutator relations from section 3.4 would also go through (adjusted to the PDE case) – but may simply be extracted from standard texts like Weinberg, since these are the standard field operators now. An important example is:

$$\left[\Phi_j(\underline{x}), \int f_n(\underline{\Phi}(\underline{y}),\underline{\Pi}(\underline{y})) d^d \underline{y}\right] = i\delta^d(\underline{x}-\underline{y}) \left(\frac{\partial f}{\partial \pi_j}(\underline{\Phi}(\underline{y}),\underline{\Pi}(\underline{y}))\right)_n \qquad (116)$$

where once again the "n" subscript refers to the normal product version of the function. The derivation should then go through as before, with some additional integration by parts.

## VI IMPLICATIONS AND POSSIBLE EXTENSIONS

These results indicate that there is an exact equivalence between the dynamics of a bosonic QFT, in the 0+1-D case, and the statistical dynamics of the corresponding "classical theory," with some minor adjustment of scalar parameters. The extension to 3+1-D should not be difficult, as just discussed.

Does this imply that bosonic QFTs are totally equivalent to classical field theory? Bell's Theorem experiments tell us that a complete equivalence is impossible. [Bell 1987, Clauser et al 1969.] However, QFT is made up of two major components, two bodies of assumptions which are combined together to generate predictions – quantum dynamics and quantum measurement. Classical field theory was not so explicit in breaking out assumptions about dynamics versus assumptions about measurement, but it, too, made strong apriori assumptions about the role of causality and statistics in measurement which were not derived at all from the assumed Lagrange-Euler equations. Therefore, it makes sense to conclude that the differences between QFT and CFT are mainly due to differences in measurement formalisms and assumptions about measurement, rather than differences in dynamics. Many authors have argued that quantum measurement could in fact be derived somehow from quantum dynamics; if true, this might imply that Lagrangian PDE models, governed by Heisenberg dynamics, would actually obey quantum measurement rules as well. Thus "classical" Lagrangian PDE models would themselves become valid candidates again for a "theory of everything."

The classical-QFT equivalence in dynamics probably extends well beyond bosonic field theory as such. LeGuillou et al [1996] claim to provide a general procedure for mapping fermionic QFTs into equivalent bosonic QFTs, based on an extension of a general procedure published by Witten in 1984. This suggests that the dynamics of the standard model itself may be equivalent to a bosonic QFT, which in turn is equivalent to a classical model.

This then suggests a possible strategy for developing a finite (not only renormalizable) variation of the standard model, by exploiting the classical-QFT equivalence in dynamics.

Many of the renormalizations required in QFT are based on effects like the infinite self-repulsion of a point charge. The same problems exist in the corresponding classical theories. But in classical field theories such effects can be eliminated by representing particles as extended bodies – more precisely, as true topological solitons. (See Makhankov et al [1994] for a review of topological solitons.)  In the electroweak part of the standard model, particles mainly derive their mass from the Higgs field, a bosonic field whose detailed properties are still almost unknown [Weinberg, 1995, chapter 21]; it would be reasonable, then, to modify the Higgs model by adding a Skyrme term, which would "explain why massive particles exist" (something closely related to explaining the mass!), and would ensure the good behavior of the corresponding classical PDE in tasks such as the prediction of scattering amplitudes. The PDE-QFT equivalence could then be exploited to prove the good behavior of the QFT. (A few auxiliary terms, analogous to Fadeev-Popov or Umezawa terms, may be needed to strictly enforce the topological constraints assumed in such models.)

This is only a possible direction for future research; it is certainly not a proof! However, please recall that when Weinberg and Salam first proposed the electroweak model in the 1960s, they did not have a proof of renormalizability – only a plausible basis for hope. The standard model is renormalizable, despite the pessimistic earlier predictions based on power-counting rules of thumb, because special properties of the model (symmetry) allow good behavior even when power-counting rules are not met. It is reasonable to hope for a similar sort of special situation here as well.

The discussion so far in this section only addresses EWT. But there is no reason to believe it could not be extended to strong nuclear interactions, the domain where Skyrme models have proven most useful in practical applications. (Witten and others presented many intriguing suggestions for how to develop more empirically-grounded models of strong interactions, using topological soliton models, in Chodos et al [1984].) It may even be possible to modify such a finite standard model, by applying the same methods John Wheeler used to convert Maxwell's Laws into an "already unified field theory." Thus there may even be some hope, in the long-term, of achieving finiteness and unification in a field theory which does not ask us to assume the existence of additional, speculative dimensions of space-time.

## References

Behrman, E.C. ,Chandrashekar, V., Wang, Z., Belur, C.K., Steck, J.E. && Skinner, S.R. [2002], "A quantum neural network computes entanglement," paper quant-ph/0202131 at xxx.lanl.gov

Bell, J.S. [1987], *The Speakable and Unspeakable in Quantum Mechanics*, Cambridge U. Press

Bohm,D. && Hiley, B. [1995] *The Undivided Universe*, Routledge

Carmichael, H. J. [1998 and 2002] *Statistical Methods in Quantum Optics*, Springer, New York, Volumes 1 and 2.

Clauser, J.F, Horne, M.A., Shimony, A. && Holt,R. A. [1969] "Proposed experiment to test local hidden-variable theories," *Phys. Rev. Lett*, **23**, p.880-884.

Chodos,A., Hadjimichael, E. && Tze, Ch., eds [1984] *Solitons in Nuclear and Elementary Particle Physics*, World Scientific, Singapore.

Coleman,S. [1975] "Quantum Sine-Gordon equation as the massive Thirring model," *Physical Review D*, **11**, p.2088-2097.

Dyson, F.J. [1949] "The S matrix in quantum electrodynamics," *Physical Review*, **75**, p.1736-1755.

Everett, H. [1973], "The theory of the universal wave function." In DeWitt, B.S. & Graham, N., eds, *The Many Worlds Interpretation of Quantum Mechanics*, Princeton U. Press.

Gershenfeld, N.A. & Chuang, I.L. [1997], "Bulk spin-resonance quantum computation," *Science*, **275**, no. 5298, p.350-6. See also Vandersypen, L.M.K., Yannoni, C.S., Sherwood, M.H. && Chuang, I.L. [1999] "Realization of effective pure states for bulk quantum computation," paper quant-ph/9905041 at xxx.lanl.gov.

Høye, J.SD., Brevik, I. && Aarseth,J.B. [2000], "The Casimir problem of spherical dielectrics: Quantum statistical and field theoretical approaches," paper quant-ph/0008088 at xxx.lanl.gov and [2001] *Physical Review E* 63 051101

Kato,T. [1995]. *Perturbation Theory for Linear Operators*, Springer.

Le Guillou, J.C., Moreno, E., Nunez, C. && Schaposnik, F.A. [1997], "Non Abelian bosonization in two and three himensions," paper hep-th 9609202 at xxx.lanl.gov and [1997] *NuclearPhysics. B,* **484 p.** 682-696

Makhankov, V., Rybakov, Y. & Sanyuk, V. [1994] *The Skyrme Model*, Springer

Mandelstam, S. [1975], "Soliton operators for the quantized sine-Gordon equation," *Physical Review D*, **11**, p.3026-3030

F.Mandl, *Introduction to Quantum Field Theory*, Wiley, 1959.

Prigogine, I.,[1994] "Mind and matter: Beyond the Cartesian dualism." In Pribram, K., ed., *Origins: Brain and Self-Organization*, Erlbaum.

Von Neumann, J. [1955], *Mathematical Foundations of Quantum Mechanics* (Beyer – tr), Princeton University Press.

Walls, D.F. & Milburn, G.F. [1994] *Quantum Optics*, Springer, New York.

Weinberg,S. [1995] *The Quantum Theory of Fields* ,Cambridge University Press, Cambridge, U.K.

Werbos, P. & Dolmatova, L., [2000], "The backwards-time interpretation of quantum mechanics – revisited with experiment," paper quant-ph 0008036 at xxx.lanl.gov.

Werbos,P. [1993], "Chaotic solitons and the foundations of physics: a potential revolution," *Applied Mathematics and Computation*, **56**, p.289-339, section IV.

Werbos, P. [1973], "An approach to the realistic explanation of quantum mechanics," *Nuovo Cimento Letters*, **29B**

Werbos, P. [1989] ,"Bell's theorem: the forgotten loophole and how to exploit it," in Kafatos, M., ed., *Bell's Theorem, Quantum Theory and Conceptions of the Universe*, Kluwer.

Witten, E. [1984], "Non-Abelian Bosonization in two dimensions," *Commun. Math. Phys.*, **92**, p. 455-472.

Yang, T. & Chua, L.O. [1999], "Implementing back-propagation-through-time learning algorithm using cellular neural networks," *Int'l J. Bifurcation and Chaos*, **9**, p.1041-1074

**Appendix. New Results**

As this paper goes to press, I have explored some extensions which cannot be presented completely here for reasons of space and time. The results here are to within scalar factors which may need adjustment to exactly match the appearance of previous equations.

An extended version of the classical density matrix may be obtained by replacing equation 73 by:

$$\widetilde{w}(\underline{\varphi},\underline{\pi}) = \exp\left(\frac{1}{\sqrt{2}}\sum_{j=1}^{n}\left((\varphi_j + i\pi_j)a_j^+ + (\varphi_j - i\pi_j)b_j^+ - (\varphi_j^2 + \pi_j^2)\right)\right)|0\rangle \quad (A.1)$$

which is a vector over the Fock space $F^{2N}(0)$, where $b_j$ and $b_j^+$ are formally defined as abbreviations for $a_{j+N}$ and $a_{j+N}^+$. This calls for the use of modified field operators; thus equations 47 and 48 are replaced by something like:

$$\Phi_j = \frac{1}{2\sqrt{2}}(a_j + a_j^+ + b_j + b_j^+) \qquad (A.2)$$

$$\Pi_j = \frac{1}{2i\sqrt{2}}(a_j - a_j^+ - b_j + b_j^+) \qquad (A.3)$$

In QFT, field operators like these are called "Schrodinger picture" field operators [Weinberg1995, Mandl 1959, Dyson 1949]. We may define classical "interaction picture" field operators by defining, for example:

$$q_j(t) = e^{-iG_0 t}\Phi_j e^{iG_0 t} \qquad (A.4)$$

$$p_j(t) = e^{-iG_0 t}\Pi_j e^{iG_0 t} \qquad (A.5)$$

where $G_0$ is:

$$\sum_j a_j a_j^+ - b_j b_j^+ \qquad (A.6)$$

These field operators (and their PDE equivalents) obey *exactly the same commutation relations over space and time* as the usual field operators of the interaction picture! This implies that a modified Schrodinger equation based on $G_0$ instead of the usual $H_0$, and with the interaction Hamiltonian defined in terms of the new field operators, would yield exactly the same Feynman integrals – in the usual S-matrix expansion – as the conventional Schrodinger equation! This representation *appears* more complicated than the equivalent conventional representation, but it actually represents something like using sin kx instead of $e^{+ikx}$, in describing the dynamics of a real field. The new dynamical matrix G is not the same as H, but since energy is still conserved, we would expect that G and H would have a complete set of joint eigenfunctions, making it legitimate to use eigenfunctions of H both in spectral calculations and in calculating the "stationary scattering states" of Heisenberg which underlie the S-matrix calculations [Dyson 1949].

These new field operators appear to have all the same essential properties that we have proven for equations 47 and 48, when applied to the extended classical density matrix. They also offer an obvious way to define the classical density matrix *in the interaction picture*.

This leads to several interesting questions. Is it possible that the definitions in the main text of this paper match standard QFT exactly in the PDE case, *but not the ODE case*, because of the infinitesimal consequences of the violations of equations 10-12? Or are there measurable differences even in the PDE case, which are overcome either by using the extended definition here, or by using the extended definition in the interaction picture? In principle, the reification methods discussed by Werbos and Dolmatova (2000) may be appropriate to the extended case here, because of the need to orthogonalize the ultimate measurements; however, the details are far beyond the scope of this paper.